\documentclass{elsart}

\usepackage{graphicx}

\usepackage{amssymb}

\usepackage{bm}

\begin{document}

\begin{frontmatter}



\title{Josephson Effect between Conventional and Rashba Superconductors}


\author[AA,BB]{Nobuhiko Hayashi},
\author[CC]{Christian Iniotakis},\,
\author[AA,BB]{Masahiko Machida},\,
\author[CC]{Manfred Sigrist}

\address[AA]{
CCSE, Japan Atomic Energy Agency, 6-9-3 Higashi-Ueno, Tokyo 110-0015, Japan
}

\address[BB]{
CREST JST, 4-1-8 Honcho, Kawaguchi, Saitama 332-0012, Japan
}

\address[CC]{
Institut f\"ur Theoretische Physik,
ETH-Z\"urich,
CH-8093 Z\"urich, Switzerland
}

\begin{abstract}
We study the Josephson effect between a conventional $s$-wave superconductor 
and a non-centrosymmetric superconductor with Rashba spin-orbit coupling.
   Rashba spin-orbit coupling affects the
Josephson pair tunneling in a characteristic way.
 The Josephson coupling can be decomposed into
 two parts, a `spin-singlet-like' and a `spin-triplet-like' component.
   The latter component can lead to shift of the Josephson phase
by $ \pi $ relative to the former coupling.
   This has important implications on interference effects and may
explain some recent
experimental results for the Al/CePt$_3$Si junction.
\end{abstract}

\begin{keyword}
Non-centrosymmetric superconductor \sep
Josephson effect \sep
Rashba spin-orbit coupling \sep
CePt$_3$Si

\PACS 74.50.+r \sep 74.20.Rp \sep 74.70.Tx
\end{keyword}
\end{frontmatter}

\section{Introduction}
 Superconductors without inversion symmetry,
the so-called non-centrosymme\-tric superconductors, have received much interest during 
recent years.
   The lack of an inversion center in the crystal lattice 
induces antisymmetric spin-orbit coupling,
leading to important modifications of the superconducting phase.
   The spin-orbit coupling displays the Rashba form
in systems such as CePt$_3$Si \cite{Bauer}
where mirror symmetry about a single plane is missing \cite{Frigeri}.
Due to such an antisymmetric spin-orbit coupling,
the Fermi surface is split into two sheets by the spin-degeneracy lifting
and the electronic spin structure on the Fermi surfaces is modified \cite{Saxena,Fujimoto}.
The specific spin structure on the Fermi surfaces due to the Rashba coupling
plays also an important role in connection with the Josephson effect, as we will show here. 
The tunneling conductance between a normal metal and
a non-centrosymmetric superconductor \cite{Yokoyama,Iniotakis,Linder}
and the Josephson effect between two non-centrosymmetric superconductors \cite{Borkje,Mandal,Borkje2}
have been investigated by various groups. 
Here we aim at the properties of 
the Josephson effect between a conventional superconductor
and a non-centrosymmetric superconductor.

 The Josephson effect between a conventional ($s$-wave spin-singlet) and
an unconventional superconductor provides a possible way to probe the spin structure
of the unconventional Cooper pairing state \cite{GL,Millis,Yip,Hasegawa,SU}.
   For the non-centrosymmetric superconductor CePt$_3$Si,
an experiment in this direction was recently performed by Sumiyama {\it et al.} \cite{Sumiyama}.
In this experiment, Al/CePt$_3$Si junctions were prepared and their Josephson effect was investigated
by applying weak magnetic fields in order to observe the interference patterns in the supercurrent. 
 A Fraunhofer-shaped pattern was observed for the tunnel junction normal to the in-plane axis of
 the tetragonal crystal of CePt$_3$Si,
while a very irregular pattern appeared for the $c$ axis tunneling \cite{Sumiyama}.
A possible explanation of these findings was recently given by Leridon {\it et al.} \cite{Varma}
proposing Cooper pairing violating time-reversal symmetry.
Here we introduce an alternative proposal which is based on 
the influence of Rashba spin-orbit coupling on the Josephson effect.
First, we derive an expression for the dc-Josephson current between
a conventional superconductor
and a non-centrosymmetric superconductor with Rashba spin-orbit coupling.
Then we analyze the specific differences between differently oriented Josephson junctions, in order to give an explanation for the observation
in the Al/CePt$_3$Si junctions.

Before going into details, we outline the basic idea. 
The Josephson current $J$ is expressed as
$J=J_c \sin \phi_{\rm ph}$, where $\phi_{\rm ph}$ is the phase difference between the
two superconductors.
 We suppose that we can decompose $J_c$ into two parts $J_c=J_1 + J_2$, where $J_1 >0$ and $J_2<0$.
 Then we assume that, for an inhomogeneous interface, the relative magnitude of the two 
contributions $ |J_1| $ and $|J_2|$ varies, such that regions with
$J_c=J_1 + J_2 >0$ and $< 0$ exist, i.e., regions with $0$- and $ \pi$-phase shifts 
$\bigl[$$J=-|J_c| \sin \phi_{\rm ph} =|J_c| \sin (\phi_{\rm ph}+\pi)$$\bigr]$, respectively.
 For such  junctions, the interference pattern would deviate strongly from the ordinary 
Fraunhofer pattern and, in particular, the central peak in the interference pattern may be missing,
as observed in other systems with random $0$- and $\pi$-junctions \cite{Hilgenkamp,Mannhart}.
   In this picture, the presence of a sufficiently strong $J_2$-component giving rise to a $\pi$ junction disturbs the Fraunhofer pattern.
   As shown later,
it follows that
$J_2 \neq 0$ for ${\hat {\bm n}} \parallel c$ and $J_2=0$ for ${\hat {\bm n}} \parallel a$,
where ${\hat {\bm n}}$ is a vector normal on the interface.
This effect would explain the experimental results \cite{Sumiyama}
that
the Fraunhofer pattern is absent (present) for the Josephson junction perpendicular to the $c$ axis
(the $a$ axis).
   In the following sections,
we will derive
an expression for the Josephson current
composed of $J_1$ and $J_2$ parts.


\section{Josephson Current}
We consider a Josephson junction between two superconductors, assuming
spherical Fermi surfaces for both superconductors, for simplicity. 
   The Fermi velocities can be written as
${\bm v}_{\mathrm F}^{L (R)}=v_{\mathrm F}^{L(R)}{\hat {\bm k}}^{L (R)}$,
where ${\hat {\bm k}}$ is the unit vector parallel to the Fermi momentum.
(We use  units with $\hbar=1$ and $k_{\mathrm B}=1$.)
The superconductors on the left- and right-hand-side of the interface are labeled by $``L"$ and $``R"$, respectively.
   According to Millis {\it et al.} \cite{Millis},
the supercurrent $J$ flowing across the interface
is given by
\begin{eqnarray}
J=N_{\mathrm F}^{L}v_{\mathrm F}^{L}
\int_{{\hat {\bm n}}\cdot {\hat {\bm k}}^{L} >0}
\frac{d \Omega_k^{L}}{4\pi}
{\hat {\bm n}}\cdot {\hat {\bm k}}^{L}
T \sum_{\omega_n} K,
\label{eq:J}
\end{eqnarray}
with
\begin{eqnarray}
K
=
\Biggl(
\frac{i}{2\pi}
\Biggr)
\frac{1}{4} \mbox{Tr}
\Bigl\{
  {\check \tau}_3
  \Bigl[
    {\check g}^{L}({\hat {\bm k}}^{L},r_\perp=0^{-},\omega_n),
    {\check S}_{R,L}^\dagger
    {\check g}^{R}({\hat {\bm k}}^{R},r_\perp=0^{+},\omega_n)
    {\check S}_{R,L}
  \Bigr]
\Bigr\},
\nonumber \\
\label{eq:K}
\end{eqnarray}
where the interface lies perpendicular to the unit vector ${\hat {\bm n}}$ located at $r_\perp=0$
and $r_\perp$ is the coordinate perpendicular to the interface.
Moreover, $N_{\mathrm F}^{L}$ is the density of states
on the Fermi surface in the left-hand-side superconductor,
and
$\omega_n=\pi T (2n+1)$ is the Matsubara frequency with $T$ as the temperature.
   The commutator 
$[{\check a},{\check b}]={\check a}{\check b}-{\check b}{\check a}$ is defined
in the usual way.
   The quasiclassical Green function ${\check g}$
is a $4\times4$ matrix
composed of blocks in $2\times2$ particle-hole space
and in $2\times2$ spin space. It can be written in the particle-hole space
as
\begin{equation}
{\check g} ({\hat {\bm k}}, r_\perp, i\omega_n) =
-i\pi
\pmatrix{
{\hat g} &
i{\hat f} \cr
-i{\hat {\bar f} } &
-{\hat {\bar g} }
}.
\label{eq:qcg}
\end{equation}
  The symbol Tr in Eq.\ (\ref{eq:K}) denotes the trace of the $4\times4$ matrix.
   The matrix ${\check \tau}_3$ is given by
\begin{equation}
{\check \tau}_3 =
\pmatrix{
{\hat \sigma}_0 &
0 \cr
0 &
-{\hat \sigma}_0
},
\end{equation}
with ${\hat \sigma}_0$ as the unit matrix in the spin space.
   The interface is characterized by the tunneling $S$ matrix ${\check S}_{R,L}$
as \cite{Millis}
\begin{equation}
{\check S}_{R,L}({\bm k}_{\parallel})
=
\pmatrix{
{\hat S}_{R,L}({\bm k}_{\parallel}) &
0 \cr
0 &
{\hat S}_{L,R}^{tr}(-{\bm k}_{\parallel})
}.
\end{equation}
 The quasiparticle momentum parallel to the interface, ${\bm k}_{\parallel}$,  
 is conserved in the tunneling process, i.e., ${\bm k}_{\parallel}^L={\bm k}_{\parallel}^R$.
 We may parametrize the $S$ matrices in the spin space
as \cite{Millis,SU}
\begin{eqnarray}
{\hat S}_{R,L}({\bm k}_{\parallel})
&=&
s_{R,L}({\bm k}_{\parallel}){\hat \sigma}_0
+
{\bm m}_{R,L}({\bm k}_{\parallel}) \cdot {\hat {\bm \sigma}},
  \\
{\hat S}_{L,R}^{tr}(-{\bm k}_{\parallel})
&=&
s_{L,R}(-{\bm k}_{\parallel}){\hat \sigma}_0
+
{\bm m}_{L,R}(-{\bm k}_{\parallel}) \cdot {\hat {\bm \sigma}}^{tr}.
\end{eqnarray}
Here we use the Pauli matrices
${\hat {\bm \sigma}}=({\hat \sigma}_x,{\hat \sigma}_y,{\hat \sigma}_z)$
for the spin space,
and $s$ and ${\bm m}$ denote
the spin-inactive and spin-active tunneling components through the interface, respectively.

For an interface potential which is invariant under time-reversal 
reflections in a plane perpendicular to the interface,
the tunneling amplitudes satisfy the relations \cite{GL,Millis}
\begin{eqnarray}
s_{R,L}({\bm k}_{\parallel})
&=&
s_{L,R}(-{\bm k}_{\parallel}),
\label{eq:s}
  \\
{\bm m}_{R,L}({\bm k}_{\parallel})
&=&
-{\bm m}_{L,R}(-{\bm k}_{\parallel})
=
c_{R,L}
{\hat {\bm n}}\times{\bm k}_{\parallel}.
\label{eq:m}
\end{eqnarray}
The $S$ matrix at an interface
between a conventional metal and a metal with Rashba spin-orbit coupling can be calculated by
applying the continuity condition for the wave function at the interface. Within this scheme,
we verified 
 the above relations 
in the limit that the difference between the volumes of the split Fermi surfaces is negligibly small, nevertheless taking account of
the modified spin structure due to Rashba spin-orbit coupling
properly \cite{haya2}. This limit allows us to perform a simple and transparent analysis of the problem keeping the essential aspects. Actually  the general expression for the $S$ matrix is rather complicated and will be discussed elsewhere \cite{haya2}.

\section{Josephson Junction between $s$-Wave and Non-Centrosymmetric Superconductors}
   We now turn to the Josephson effect 
between a spin-singlet $s$-wave superconductor (left-hand side) and
a non-centrosymmetric one (right-hand side) with Rashba spin-orbit coupling
$\sim {\bm \lambda} \cdot {\hat {\bm \sigma}}$,
where ${\bm \lambda}=(-k_y^R, k_x^R, 0)$ \cite{haya}.
  The qualitative properties of such a junction can be obtained utilizing
bulk Green functions as in Refs.\ \cite{GL,Millis,Hasegawa,SU}. 
In the spin-singlet $s$-wave superconductor with  $\Psi_{s}$ as the order parameter,
the Green functions  are given as \cite{Klein}
\begin{eqnarray}
{\hat g}^{L}
&=&
\frac{\omega_n {\hat \sigma}_0}{\sqrt{\omega_n^2+|\Psi_{s}|^2}},
\qquad
{\hat {\bar g} }^{L}
=
\frac{\omega_n {\hat \sigma}_0}{\sqrt{\omega_n^2+|\Psi_{s}|^2}},
\nonumber  \\
{\hat f}^{L}
&=&
\frac{\Psi_{s} i {\hat \sigma}_y }{\sqrt{\omega_n^2+|\Psi_{s}|^2}},
\qquad
{\hat {\bar f} }^{L}
=
\frac{\Psi_{s}^* (-i {\hat \sigma}_y) }{\sqrt{\omega_n^2+|\Psi_{s}|^2}}.
\end{eqnarray}
On the non-centrosymmetric superconductor side, they
   have the form
$\bigl[$${\bar k}'_\pm = {\bar k}_y^R \pm i{\bar k}_x^R$, 
$({\bar k}_x^R)^2+({\bar k}_y^R)^2=1$$\bigr]$ \cite{haya}:
\begin{eqnarray}
{\hat g}^{R}
=
\pmatrix{
 g_{+} &
 -{\bar k}'_+ g_{-}  \cr
 -{\bar k}'_- g_{-}  &
 g_{+}
},
\quad
{\hat {\bar g} }^{R}
=
\pmatrix{
 g_{+} &
 {\bar k}'_- g_{-}  \cr
 {\bar k}'_+ g_{-}  &
 g_{+}
},
\nonumber \\
{\hat f}^{R}
=
\pmatrix{
 {\bar k}'_+ f_{-}  &
 f_{+}  \cr
 -f_{+} &
 -{\bar k}'_- f_{-}
},
\quad
{\hat {\bar f} }^{R}
=
\pmatrix{
 {\bar k}'_- {\bar f}_{-}  &
 -{\bar f}_{+} \cr
 {\bar f}_{+}  &
 -{\bar k}'_+ {\bar f}_{-}
}.
\label{eq:G}
\end{eqnarray}
We use the notation,
$g_\pm=(g_{\rm I} \pm g_{\rm II})/{2}$,
$f_\pm=(f_{\rm I} \pm f_{\rm II})/{2}$,
${\bar f}_\pm=({\bar f}_{\rm I} \pm {\bar f}_{\rm II})/{2}$,
$g_{\rm I,II}
=
    { \omega_n }/{B_{\rm I,II}}$,
$f_{\rm I,II}
=
    { \Delta_{\rm I,II} }/{B_{\rm I,II}}$,
${\bar f}_{\rm I,II}
=
{ \Delta_{\rm I,II}^* }/{B_{\rm I,II}}$,
$B_{\rm I,II}
=
\sqrt{
      \omega_n^2 
      +|\Delta_{\rm I,II}|^2
    }$,
$\Delta_{\rm I} = \Psi+\Delta\sin\theta_R$,
and
$\Delta_{\rm II} = \Psi-\Delta\sin\theta_R$.
The superconducting order parameters $\Delta_{\rm I,II}$ are defined 
on the split Fermi surfaces I and II,
and $\Psi$ ($\Delta$) stands for the singlet (triplet) order-parameter component
in the non-centrosymmetric superconductor. It is important that 
the relative phase between $\Psi$ and $\Delta$ is 0 or $\pi$ \cite{haya}.
Furthermore, $\theta_R$ denotes  the angle relative to the $k_z^R$ axis with
$\sin\theta_R \geq 0$.

The above Green functions can be inserted into Eq.\ (\ref{eq:K})
for $K$.  Using Eqs.\ (\ref{eq:s}) and (\ref{eq:m}),
we finally arrive at the following result,  
\begin{eqnarray}
K
&=&
\frac{\pi(1+\delta)}{ D_{\rm I} }
\mbox{Im}
\Bigl\{
w_0^*
\Psi_{s}
   \Delta_{{\rm I}}^*
\Bigr\}
\nonumber  \\
& & { }
+
\frac{\pi(1-\delta)}{ D_{\rm II} }
\mbox{Im}
\Bigl\{
w_0^*
\Psi_{s}
   \Delta_{{\rm II}}^*
\Bigr\}
\nonumber  \\
& & { }
+
\frac{\pi(1+\delta)}{ D_{\rm I} }
\mbox{Im}
\Bigl\{
{\bar {\bm \lambda}} \cdot {\bm w}^*
\Psi_{s}
 \Delta_{{\rm I}}^*
\Bigr\}
\nonumber  \\
& & { }
+
\frac{-\pi(1-\delta)}{ D_{\rm II} }
\mbox{Im}
\Bigl\{
{\bar {\bm \lambda}} \cdot {\bm w}^*
\Psi_{s}
 \Delta_{{\rm II}}^*
\Bigr\},
\label{eq:K2}
\end{eqnarray}
where
${\bar {\bm \lambda}}=(-{\bar k}_y^R, {\bar k}_x^R, 0)$ and
$D_{\rm I,II}=2 \sqrt{\omega_n^2+|\Psi_{s}|^2}
              \sqrt{\omega_n^2+|\Delta_{{\rm I,II}}|^2} $.
 The split Fermi surfaces (I and II) are taken into account also by the             
parameter $\delta$ which denotes
the difference of the density of states. 
$\bigl[$$\delta=(N_{\rm I}-N_{\rm II})/2N_0$, $2N_0=N_{\rm I}+N_{\rm II}$$\bigr]$~\cite{haya}.
The interface is described by means of the parameters, the scalar $w_0$
and the vector  ${\bm w}=(w_x, w_y, w_z)$, given by
\begin{eqnarray}
w_0
=
|s_{R,L}|^2
+|c_{R,L}|^2   \bigl|{\hat {\bm n}}\times{\bm k}_{\parallel} \bigr|^2,
\label{eq:w0}
\end{eqnarray}
\begin{eqnarray}
{\bm w}
=
2\mbox{Re}
\{
s_{R,L} c_{R,L}^*
\}
{\hat {\bm n}}\times{\bm k}_{\parallel}.
\label{eq:wvec}
\end{eqnarray}

Within this scheme, we find that  the first and second terms in Eq.\ (\ref{eq:K2}) 
are similar to the Josephson coupling between spin-singlet superconductors
(``singlet-like" coupling),
while the third and fourth terms resemble the coupling
between spin-singlet and spin-triplet superconductors 
(``triplet-like" coupling) \cite{Millis,SU}.
This decomposition is independent of
the explicit mixing of singlet-triplet pairing in the non-centrosymmetric superconductor,
because
Eq.\ (\ref{eq:K2}) has the same form even in the ``pure singlet" ($\Delta=0$)
and ``pure triplet" ($\Psi=0$) pairing cases.
   The origin of the factor
${\bar {\bm \lambda}} \cdot {\bm w}^*$ in Eq.\ (\ref{eq:K2})
is {\it not} the triplet pairing, but is the characteristic spin structure on the Fermi surfaces due to
Rashba spin-orbit coupling $\sim {\bm \lambda} \cdot {\hat {\bm \sigma}}$.

\section{Directional dependence of Josephson current}
   The Josephson current is obtained by integrating
Eq.\ (\ref{eq:K2}) over the Fermi surface
($\sim \int_{{\hat {\bm n}}\cdot{\hat {\bm k}} >0} d\Omega_k ({\hat {\bm n}}\cdot{\hat {\bm k}}) K
= \int_{{\hat {\bm n}}\cdot{\hat {\bm k}} >0} d\phi \d\theta\sin\theta ({\hat {\bm n}}\cdot{\hat {\bm k}}) K$).
   In the first and second terms of Eq.\ (\ref{eq:K2})
(singlet-like coupling),
$w_0$ is always positive according to Eq.\ (\ref{eq:w0}),
and therefore these terms correspond to $J_1 \sin \phi_{\rm ph}$ ($J_1>0$).
Note that
$w_0$ is real [Eq.\ (\ref{eq:w0})] and
the factor $\mbox{Im}
\bigl\{
w_0^*
\Psi_{s}
   \Delta_{{\rm I,II}}^*
\bigr\}$
in Eq.\ (\ref{eq:K2})
leads to $w_0|\Psi_{s}| |\Delta_{{\rm I,II}}| \sin \phi_{\rm ph}$.
   On the other hand,
the third and fourth terms of Eq.\ (\ref{eq:K2}) 
(triplet-like coupling)
can be negative
because of the coefficient $c_{R,L}$ of the spin-active tunneling $\bigl[$Eq.\ (\ref{eq:m})$\bigr]$,
the sign of which depends on an interface potential formed
at the junction between two different materials \cite{haya2}.
The actual interface potential formed between Al and CePt$_3$Si
is unknown. However, we assume here that the interface potential gives rise to a negative sign
for $c_{R,L}$.
 Under this assumption, the third and fourth terms of Eq.\ (\ref{eq:K2}) can correspond
to $J_2 \sin \phi_{\rm ph}$ ($J_2<0$). Note that from Eq.\ (\ref{eq:wvec}) it follows that
${\bm w}$ is real, and
the factor $\mbox{Im}
\bigl\{
{\bar {\bm \lambda}} \cdot {\bm w}^*
\Psi_{s}
   \Delta_{{\rm I,II}}^*
\bigr\}$
in Eq.\ (\ref{eq:K2})
leads to ${\bar {\bm \lambda}} \cdot {\bm w} |\Psi_{s}| |\Delta_{{\rm I,II}}| \sin \phi_{\rm ph}$.
Now, we will demonstrate that $J_2$, namely the integration of
the third and fourth terms in Eq.\ (\ref{eq:K2}), becomes zero
for a certain direction of the interface.

The interface (${\hat {\bm n}} \parallel a$) normal to the $x$ axis ($a$ axis) yields
${\bm w} \sim {\hat {\bm n}}\times{\bm k}_{\parallel} \sim (0,-\cos\theta,\sin\phi\sin\theta)$,
where we use spherical coordinates according to
${\hat {\bm k}}=(\cos\phi\sin\theta, \sin\phi\sin\theta, \cos\theta)$
and $({\bar k}_x^R, {\bar k}_y^R)=(\cos\phi, \sin\phi)$.   
Hence we obtain the factor
$
{\bar {\bm \lambda}} \cdot {\bm w}
=
(-{\bar k}_y^R) w_x
+
{\bar k}_x^R w_y
\sim
-\cos\phi \cos\theta
$, which  is an {\it odd} function with respect to $\theta'$ ($=\theta-\pi/2$).
   All other factors, $D_{\rm I,II}$ and $\Delta_{\rm I,II}$ in Eq.\ (\ref{eq:K2}),
keep their sign with respect to $\theta$ ($0 \leq \theta \leq \pi$).
From these, we conclude that
$\int_{{\hat {\bm n}}\cdot{\hat {\bm k}} >0} d\phi \d\theta\sin\theta
({\hat {\bm n}}\cdot{\hat {\bm k}})
\{\mbox{3rd and 4th terms of }K\}
=0$ for ${\hat {\bm n}} \parallel a$,
because of the factor $\cos\theta$
originating from
${\bar {\bm \lambda}} \cdot {\bm w}$,
where the integral range for $\theta$ is $0 \leq \theta \leq \pi$.

In contrast,  for the interface (${\hat {\bm n}} \parallel c$) normal to the $z$ axis ($c$ axis),
${\bm w} \sim {{\hat {\bm n}}\times{\bm k}_{\parallel}} \sim
(-\sin\phi\sin\theta, \cos\phi\sin\theta, 0)$, and we get 
${\bar {\bm \lambda}} \cdot {\bm w}
=
(-{\bar k}_y^R) w_x
+
{\bar k}_x^R w_y
\sim
(\sin^2\phi + \cos^2\phi) \sin\theta
=\sin\theta
$.
This factor remains
{\it positive} over the integral range $0 \leq \theta \leq \pi/2$.
The other factors such as $D_{\rm I,II}$ and $\Delta_{\rm I,II}$ in Eq.\ (\ref{eq:K2})
show no sign change in the whole range of $\theta$ ($0 \leq \theta \leq \pi/2$).
Thus we find
$\int_{{\hat {\bm n}}\cdot{\hat {\bm k}} >0} d\phi \d\theta\sin\theta
({\hat {\bm n}}\cdot{\hat {\bm k}})
\{\mbox{3rd and 4th terms of }K\}
\neq 0$ for ${\hat {\bm n}} \parallel c$,
because the integrand does not change its sign over the integral range.

Consequently, the results $J_2 \neq 0$ for ${\hat {\bm n}} \parallel c$ and $J_2=0$ for ${\hat {\bm n}} \parallel a$,
have been obtained. They explain the experimental results
for the Al/CePt$_3$Si junctions \cite{Sumiyama}
as discussed in Introduction.
   From a symmetry point of view, our discussion would not be changed qualitatively for
the case of anisotropic Fermi surfaces. Symmetries such as the mirror symmetry about the $k_x k_y$ plane remain valid, 
even under the influence of antisymmetric (Rashba) spin-orbit coupling \cite{Frigeri2}.
For this reason, the above argumentation on the signs of the integrands must hold even 
for anisotropic Fermi surfaces.

\section{Conclusion}
   We investigated the behavior of a Josephson junction between a singlet $s$-wave superconductor
and a non-centrosymmetric superconductor with Rashba spin-orbit coupling.
   The expression for the Josephson current-phase relation was derived for such a junction.
   This allowed us to give a possible  explanation for
the recent experimental results \cite{Sumiyama} for the Al/CePt$_3$Si junction.
   Furthermore, we anticipate that in the absence of an external magnetic field,
spontaneous magnetic fluxes could appear along the interface normal to the $c$ axis
of CePt$_3$Si owing to random $\pi$- and 0-junctions,
which can be observed experimentally, in principle, by scanning SQUID microscopes
as in the case presented in Ref.\ \cite{Mannhart}. 
Moreover, Andreev bound states can be formed at surfaces of certain orientations in 
non-centrosymmetric superconductors  \cite{Iniotakis,Borkje2}.
The influences of such bound states have been neglected here for the qualitative discussions
as in Refs.\ \cite{GL,Millis,Hasegawa,SU}.
A more detailed analysis  taking these aspects into account is left for future studies.

\section*{ACKNOWLEDGMENTS}
We are grateful to
A.\ Sumiyama,
D.\ F.\ Agterberg, P.\ A.\ Frigeri, S.\ Fujimoto
and K.\ Wakabayashi
for helpful discussions.
We also acknowledge financial support
from the Swiss Nationalfonds and the NCCR MaNEP.




\begin{thebibliography}{00}

\bibitem{Bauer}
E. Bauer,
G. Hilscher, H. Michor, Ch. Paul, E. W. Scheidt, A. Gribanov,
Yu. Seropegin, H. No\"el, M. Sigrist, P. Rogl,
Phys. Rev. Lett. 92 (2004) 027003.

\bibitem{Frigeri}
P. A. Frigeri, D. F. Agterberg, A. Koga, and M. Sigrist,
Phys. Rev. Lett. 92 (2004) 097001;
and references therein.

\bibitem{Saxena}
S. S. Saxena, P. Monthoux,
Nature 427 (2004) 799.

\bibitem{Fujimoto}
S. Fujimoto,
J. Phys. Soc. Jpn. 76 (2007) 051008, cond-mat/0702585.

\bibitem{Yokoyama}
T. Yokoyama, Y. Tanaka, J. Inoue,
Phys. Rev. B 72 (2005) 220504; \\
Phys. Rev. B 74 (2006) 035318.

\bibitem{Iniotakis}
C. Iniotakis, N. Hayashi, Y. Sawa, T. Yokoyama, U. May, Y. Tanaka, M. Sigrist,
Phys. Rev. B 76 (2007) 012501.

\bibitem{Linder}
J. Linder, A. Sudb{\o},
Phys. Rev. B 76 (2007) 054511.

\bibitem{Borkje}
K. B{\o}rkje, A. Sudb{\o},
Phys. Rev. B 74 (2006) 054506.

\bibitem{Mandal}
S. S. Mandal, S. P. Mukherjee,
J. Phys.: Condens. Matter 18 (2006) L593.

\bibitem{Borkje2}
K. B{\o}rkje,
Phys. Rev. B 76 (2007) 184513.

\bibitem{GL}
V. B. Geshkenbein, A. I. Larkin,
JETP Lett 43 (1986) 395; \\
J. A. Sauls, Z. Zou, P. W. Anderson,
unpublished.

\bibitem{Millis}
A. Millis, D. Rainer, J. A. Sauls,
Phys. Rev. B 38 (1988) 4504.

\bibitem{Yip}
S.-K. Yip, O. F. De Alcantara Bonfim, P. Kumar,
Phys. Rev. B 41 (1990) 11214.

\bibitem{Hasegawa}
Y. Hasegawa,
J. Phys. Soc. Jpn. 67 (1998) 3699.

\bibitem{SU}
M. Sigrist, K. Ueda,
Rev. Mod. Phys. 63 (1991) 239,
Sec. IV B.

\bibitem{Sumiyama}
A. Sumiyama, K. Nakatsuji, Y. Tsuji, Y. Oda, T. Yasuda, R. Settai, Y. Onuki,
J. Phys. Soc. Jpn. 74 (2005) 3041.

\bibitem{Varma}
B. Leridon, T.-K. Ng, C. M. Varma,
Phys. Rev. Lett. 99 (2007) 027002.

\bibitem{Hilgenkamp}
H. Hilgenkamp, J. Mannhart, B. Mayer,
Phys. Rev. B 53 (1996) 14586.

\bibitem{Mannhart}
J. Mannhart, H. Hilgenkamp, B. Mayer, Ch. Gerber, J. R. Kirtley, K. A. Moler, M. Sigrist,
Phys. Rev. Lett. 77 (1996) 2782.

\bibitem{haya2}
N. Hayashi, et al.,
unpublished.





\bibitem{haya}
N. Hayashi, K. Wakabayashi, P. A. Frigeri, M. Sigrist,
Phys. Rev. B 73 (2006) 024504;
Phys. Rev. B 73 (2006) 092508.

\bibitem{Klein}
U. Klein,
J. Low Temp. Phys. 69 (1987) 1.








\bibitem{Frigeri2}
P. A. Frigeri, D. F. Agterberg, M. Sigrist,
New J. Phys. 6 (2004) 115.





%

%

















%











\end{thebibliography}
\end{document}